\documentclass[iop]{emulateapj}

\usepackage{amsmath}
\usepackage{lipsum}
\usepackage{graphicx}
\usepackage[caption=false]{subfig}

\slugcomment{Draft Version 11/13/2017}

\shorttitle{Deep Water Cycling and Delayed Onset Cooling of the Earth}
\shortauthors{Seales and Lenardic}

\begin{document}


\title{Deep Water Cycling and Delayed Onset Cooling of the Earth}


\author{Johnny Seales\altaffilmark{1} and Adrian Lenardic\altaffilmark{1}}
\affil{Department of Earth, Environment and Planetary Sciences, Rice University,
    Houston, TX 77005}
\email{jds16@rice.edu}
\email{ajns@rice.edu}



\begin{abstract}
Changes that occur on our planet can be tracked back to one of two energy sources: the sun and the Earth's internal energy. The motion of tectonic plates, volcanism, mountain building and the reshaping of our planet's surface over geologic time depend on the Earth's internal energy. Tectonic activity is driven by internal energy and affects the rate at which energy is tapped, i.e., the cooling rate of our planet. Petrologic data indicate that cooling did not occur at a constant rate over geologic history. Interior cooling was mild until ~2.5 billion years ago and then increased (Figure 1). As the Earth cools, it cycles water between its rocky interior (crust and mantle) and its surface. Water affects the viscosity of mantle rock, which affects the pace of tectonics and, by association, Earth cooling. We present suites of thermal-tectonic history models, coupled to deep water cycling, to show that the petrologically constrained change in the Earth's cooling rate can be accounted for by variations in deep water cycling over geologic time. The change in cooling rate does not require a change in the global tectonic mode of the Earth. It can be accounted for by a change in the balance of water cycling between the Earth's interior and its surface envelopes. The nature and timing of that water cycling change can be correlated to a change in the nature of continental crust and an associated rise of atmospheric oxygen. The prediction that the rise of oxygen should then be correlated, in time, to the change in the Earth's cooling rate is consistent with data constraints.
\end{abstract}



\keywords{thermal history; deep water cycle}

\section*{ }Solid planet cooling, and how it connects to volcanic-tectonic evolution, defines the Earth's thermal history. At present, the cooling of our planet's interior is associated with plate tectonics. At mid-ocean ridges, plates are created by decompression melting of mantle rock. As plates spread, they cool and transfer heat. Oceanic plates eventually subduct back into the mantle, and cold rock is mixed into the progressively cooling interior of our planet. Heat transfer associated with macroscopic motion is, by definition, thermal convection. Plate tectonics is thus a component of mantle convection and the Earth's convective cooling. Given this, a change in cooling rate (Figure 1) could reasonably be taken to indicate a change in the tectonic mode of our planet [Condie et al., 2016]. There is an alternative that does not require a change in the global tectonic mode of the Earth. The alternative hypothesis is connected to the cycling of water between Earth's interior and surface.\\

Figure 2 shows petrological constraints [Condie et al., 2016; Herzberg et al., 2010] along with a thermal history model that couples thermal convection in a planet's interior to deep water cycling [Sandu et. al., 2011]. This is one model case from thousands we have run (Figures 3 and 4). Our modeling suites account for temperature and hydration effects on mantle viscosity [Kohlstedt, 2006; Li et al, 2008], melting and volcanism that can dehydrate the mantle [Hirschmann, 2000; Katz et. al., 2003], and processes that recycle surface water to the mantle [Ulmer and Trommsdorff, 1995; Rupke et al., 2004]. Mantle de-watering occurs at mid ocean ridges and re-watering at subduction zones. The amount of water initially in the mantle is an initial condition. The fraction of water in melt that makes it to the surface, $(\chi_d)$, and the fraction of water carried into the deep mantle with subducting slabs, $(\chi_r)$, represent de- and re-watering efficiencies. Surface heat flow follows the scaling form of $Nu\sim Ra^{\beta}$, where $Nu$ is the nondimensional heat flux, $Ra$ quantifies the vigor of mantle convection, and $\beta$ is a scaling parameter that accounts for variations in plate strength affecting the motion of tectonic plates and associated mantle cooling [Conrad and Hager, 1999; Korenaga, 2003; Hoink et al., 2013]. In all cases, plate velocity parameterizations are such that model plate velocities, after 4.6 Gy of evolution, match present day values. Further model description can be found in the supplemental material.\\

\begin{figure}
  \centering
  \includegraphics[width=\linewidth]{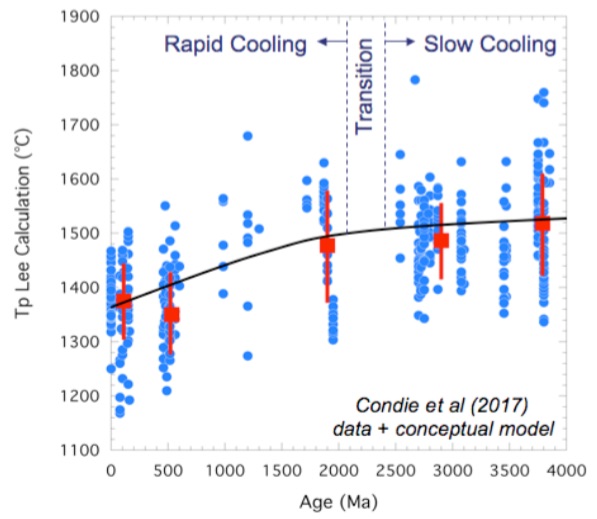}
  \caption{Figure 1: Petrological data from Condie et. al. [2016] along with their conceptual interpretation of the data.}
\end{figure}

The coupling between deep water cycling and thermal convection in the Earth's mantle leads to feedbacks that affect planetary evolution [Crowley et. al., 2011]. Internal temperature changes alter the nature of deep water cycling, which feeds back and alters mantle thermal evolution. If, for example, mantle temperature is relatively high, melt production can increase. Accompanying higher melt production are greater rates of mantle de-watering, leading to a stiffer mantle [Kohlstedt, 2006; Li et al, 2008] and a less efficient heat transfer. This, in turn, can act to increase mantle temperatures. Heating proceeds until temperature effects on mantle viscosity can outpace those of mantle dehydration, thereby reducing viscosity and allowing for faster convection, which acts to offset the heating effect associated with dehydration. Recycling water into the Earth's interior is also associated with potential feedbacks. For cooler conditions, hydrated mineral phases are stable to greater depth, [Ulmer and Trommsdorff, 1995; Rupke et al., 2004] and as a result, the ability to cycle water into the interior can increase with mantle cooling. Water cycled into the mantle lowers mantle viscosity, which decreases the resistance to plate overturn and, by association, increases mantle cooling. The strength of these coupled feedbacks evolve through time, and this opens the potential of a wider diversity of thermal histories as compared to a situation without deep water cycling. \\

The model of Figure 2 started warm, contained 1.5 present day ocean mass equivalents (OM) of water in the mantle and evolved as water cycled between the surface and interior. The evolution consisted of three phases: rapid initial cooling, near constant temperature, and a steady decrease to present day. The first phase was facilitated by a hot and hydrated mantle. These factors decrease mantle viscosity which increased convective vigor and mantle cooling. The second phase is highlighted by a flattening of the cooling curve consistent with the data of Condie et al. [2016]. The data of Herzberg et al. [2010] is associated with warmer temperatures but overall trends are similar. Mantle degassing continued beyond the initial rapid phase of mantle cooling. Continued mantle dehydration drove an increase in viscosity which damped the efficiency of heat transport. Decreased heat transport drove an increase in mantle temperature which offset the tendency for the mantle temperature to drop as internal heat sources decayed. This occurred to the point where there was a near balance between the two effects which lead to a flat line cooling phase. The flat line behavior correlated with a model water evolution phase during which surface water mass increased due to the dominance of outgassing from the interior (Figure 2b). Over time, the rate of mantle dehyrdration waned and could not maintain a flat line trend. As the mantle cooled, lithospheric plates thickened and the stability field for hydrated minerals deepened allowing a larger capacity of water rich rock to be transported back into the deep mantle (subduction zones transitioned from being dry to wet). At $\sim$2.0 Gyr of model time, mantle de- and re-hyrdration became comparable which terminated the flat line cooling phase. The final phase was characterized by internal temperature tracking the decay of radiogenic heat sources in the mantle. For a number of cases explored, rehydration exceeded mantle dehydration which lead to a steeper final cooling phase. \\

\begin{figure}[b!]
  \includegraphics[width=\linewidth]{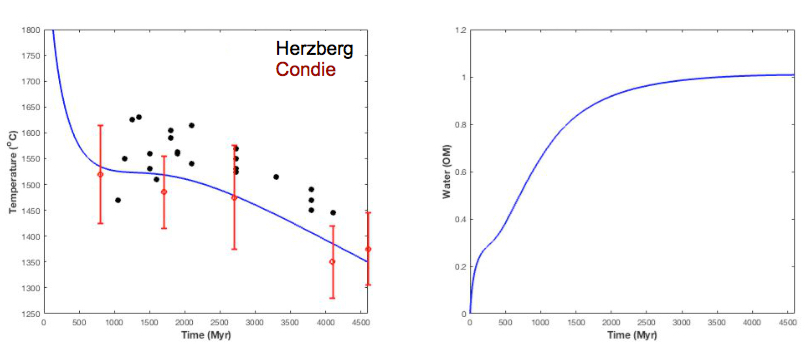}
  \caption*{Figure 2: Representative thermal evolution with deep water cycling effects compared to petrologic data (a). The blue curve is the thermal evolution model. The red and black dots are data from Condie et. al. [2016] and Herzberg et. al. [2010]. (b) Model surface water evolution. }
  \label{fig:appfig1}
\end{figure}

\begin{figure*}[t]
  \includegraphics[width=\linewidth]{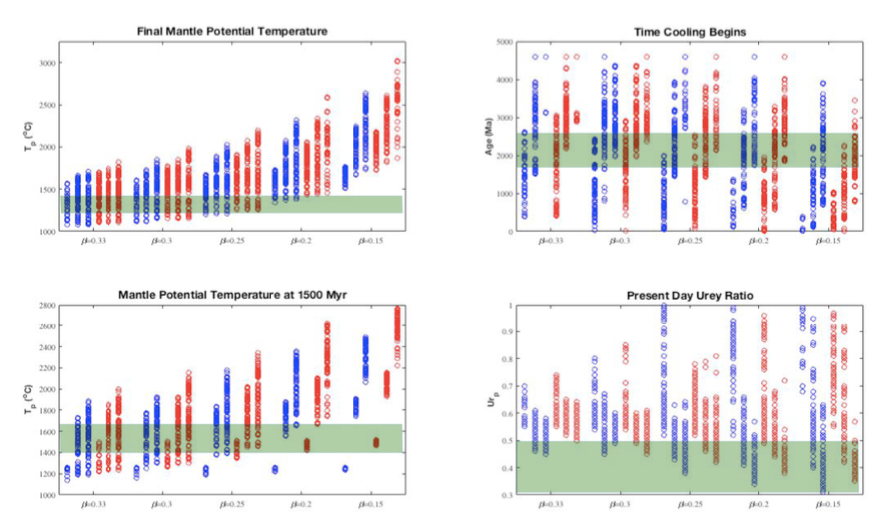}
  \caption*{Figure 3: Four panels representing the final mantle potential temperature (a), the onset time of the most recent phase of mantle cooling (b), the mantle temperature after 1.5 Gy of model evolution (c), and the present day Urey ratio (d).  }
  \label{fig:appfig1}
\end{figure*}

Figure 2 plots an illustrative case from a large number of models that were run over varied model assumptions, parameter values, and initial conditions. In Figure 3, each open circle represents the result of one model for a distinct combination of initial conditions and parameter values. The four plots represent the final mantle temperature, the age at which the most recent cooling phase initiated, the mantle temperature after 1.5 Gy of model evolution, and the present day ratio of mantle heat sources to mantle heat loss (referred to as the mantle Urey ratio). The Urey ratio values account for the effects of present day continental area [Grigne et al., 2001; Lenardic et al., 2011]. There are six column sets of model cases for each value of $\beta$. Decreasing $\beta$, from left to right, parameterizes the effects of plate strength progressively offering enhanced resistance to plate motion [Conrad and Hager, 1999]. The three red column sets, for each fixed $\beta$ suite, represent an initial mantle heat source density that is chondritic [e.g., Schubert et al., 1980] while the blue sets represent reduced mantle heat source density [Jackson and Jellinek, 2013]. The three different columns, for each fixed $\beta$ and heat source density sub-suite, represent increasing initial thermal conditions (from right to left columns, the starting temperature is 900 K, 1600 K, and 2300 degrees). Each individual model set column uses the results of 128 cases that varied initial water volume along with de- and re-gassing parameters. Initial water volumes ranged from 0.1 to 6 ocean mass equivalents. Re- and de-gassing efficiency parameters spanned the high to low parameter ranges from Sandu et al. [2011]. \\

Figure 3 demonstrates the diversity of model behavior associated with feedbacks between deep water cycling and thermo-tectonic evolution. Differing combinations of de- and re-watering efficiencies, together with variable thermal parameters and initial conditions, can produce a range of thermal histories. The green transparent rectangles in Figure 3 delineate the range of model cases that are consistent with data constraints. Petrological data provide a constraint for onset time of the Earth's most recent cooling phase and mantle potential temperatures in the flat line cooling phase. Although the absolutes for different petrologic data sets are not in exact agreement, the trends match reasonably well (figure 2a). In particular, both data sets indicate that from the onset of the most recent cooling phase to the present day, the mantle has cooled by 100-200 $^o$C. Another constraint on thermal evolution is the present day Urey ratio (Ur) which, within data uncertainty, is between 0.2 and 0.5 [Jaupart et al., 2007]. \\ 

Grouping model results into variable $\beta$ sets reflects the fact that different values of  $\beta$ are associated with different hypothesis regarding solid planet dynamics. A model with $\beta=0.33$ is based on the hypothesis that the dominant resistance to the motion of tectonic plates comes from interior mantle viscosity [Davies 1980; Schubert et al., 1980]. A model that with $\beta=0$ is based on the hypothesis that the dominant resistance to the motion of tectonic plates comes from the strength of plates themselves with any changes in internal mantle viscosity having no effect on plate velocity [Christensen, 1984; 1985]. Intermediate values reflect variable hypotheses regarding the balance between internal mantle and plate sourced resistance to plate motion [Conrad and Hager, 1999]. Our modeling strategy follows that of McNamara and VanKeken [2000]: running a large number of models, under varied assumptions, allows different hypotheses to be assessed against each other in a statistical manner. It is no surprise, given parameter and initial condition uncertainties, that cases can be found that match data constraints from models that are based on fundamentally different physical assumptions. Given this, a probabilistic approach becomes necessary to: 1) determine if the ability of any hypothesis to match data is statistically meaningful (that is, to determine if model results, that match data, may be outliers in parameter space), and 2) to discriminate between competing hypothesis. \\

The Urey ratio has come to be seen as a key value that can be used to discriminate between competing thermal evolution hypotheses [Conrad and Hager, 1999; Korenaga, 2003; Jaupart et al., 2007]. Model Urey ratio values (Figure 3d) can be viewed probabilistically as in Figure 4. Model output distribution, when all $\beta$ cases are plotted together, is not uni-modal. The model Ur peak at $\sim$0.6 is close to that obtained by first generation thermal history models that did not account for deep water cycling and/or the effects of strong plates resisting motion [Davies 1980; Schubert et al., 1980]. The peak at $\sim$0.35  is more in line with data constraints [Jaupart et al., 2007]. \\

\begin{figure*}
  \includegraphics[width=\linewidth]{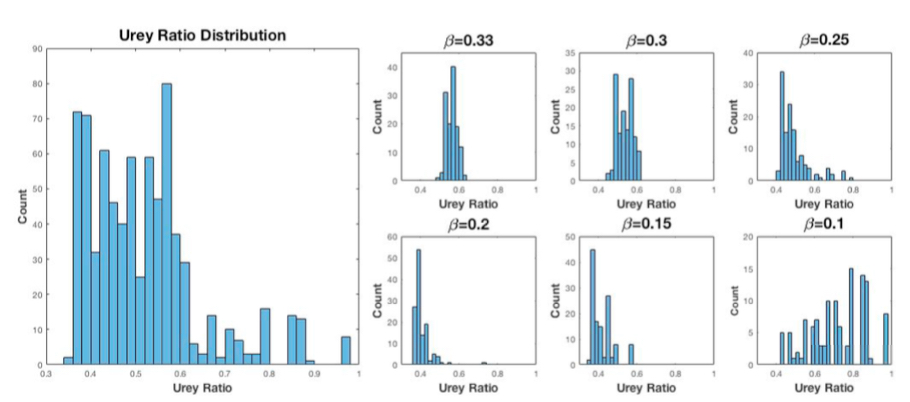}
  \includegraphics[width=\linewidth]{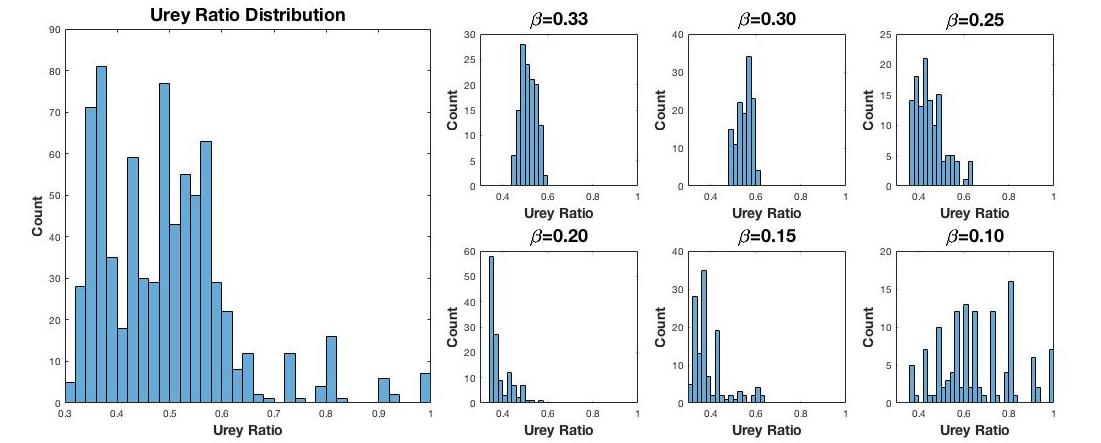}
  \caption*{Figure 4: Distribution of present day Urey ratios. The top set of panels is for cases with chordritic heat sources [e.g., Schubert et al., 1980]. The bottom set is for cases with reduced mantle heat source density [Jackson and Jellinek, 2013]. The large plot at the left, for the upper and lower sets, shows results for all $\beta$ values. The smaller plots to the right of each large plot show the individual $\beta$ value distributions.}
  \label{fig:appfig1}
\end{figure*}

Including deep water cycling into classic thermal history models that do not include the effects of strong plates can lower Urey ratio values to a point such that the data constraints can be matched. This was already noted by Sandu et al. [2011]. A probabilistic approach, which covers larger parameter space allowing distribution functions to be constructed, shows that although this conclusion is valid, the model distribution peak is located outside the data window (Figure 4, top panels). Non-classical assumptions regarding initial heat source density can bring the distribution peak within the data window (Figure 4, bottom panel). None the less, models that allow plate strength to provide a component of resistance to plate motion are statistically preferred as they bring the distribution peak deeper into the data window (Figure 4, smaller panels). As $\beta$ decreases, the model Ur distribution peak shifts towards lower values. When $\beta$ is equal to 0.20, the model distribution becomes uni-modal with a peak Urey ratio of $\sim$0.35 (Figure 4). However, there is a limit to this behavior. As $\beta$ is lowered towards zero, model output looses a uni-modal character and spans a wider Urey ratio range. \\

Shifting $\beta$ towards zero means that plate velocities progressively do not scale with convective vigor and as such, become constant over the full evolution time. Any enhanced cooling potential, due to enhanced subduction under hotter mantle conditions, is thus damped. This weakens a negative, i.e., buffering, feedback within the system. Weakening this negative feedback means that final model results become more sensitive to initial conditions [Korenaga, 2016]. If the negative feedback is removed altogether, $\beta \leq 0$, then the model system has no stable attracting state and results become highly sensitive to assumed initial conditions and the effects of parameter uncertainties become large [Moore and Lenardic, 2015]. This is reflected in Figure 4  for the lowest $\beta$ suites which are characterized by model outputs with a multi-modal distribution. Removing a thermal feedback also alters water cycling. Cooler conditions allow for enhanced water cycling into the mantle (water cycling into the mantle is also a cooling feedback as it lowers viscosity which allows for enhanced convective overturn and heat loss). The lack of a subduction cooling feedback can move the system towards one that is dominated by mantle de-watering over larger portions of model evolution. Thus, de-watering can outweigh re-watering over a larger portion of potential parameter space. This tends to shift cooling times closer toward present day and can also lead to model distributions that allow for higher Urey ratio values [Crowley et al., 2011]. \\

Many of the model cases explored do not match observational constraints for the Earth. However, the range of potential solutions that can match data constraints, given uncertainties in the data, initial conditions, model parameter values, and variable hypotheses regarding the role of plate strength, is significant (the percentage breakdown of models that can match data constraints is provided in the supplemental material). A very low $\beta$ hypothesis is associated with an extreme sensitivity to parameter uncertainty. In addition, the lowest $\beta$ cases all run too hot relative to data constraints (Figure 3). Collectively, this makes a low $\beta$ hypothesis statistically weak and cases with a $\beta$ range between 0.33-0.20 are favored. Within that model range, an initial heat source distribution lower than chondritic is also favored, whereas very cool starting conditions are not. Of the models subject to those statistical preferences, 13.3 percent match data constraints (the total number of models in that subset still exceeds 1000 cases). That is to say, solutions of the type shown in Figure 2 are not outliers. From this statistical perspective, petrologic data trends can be accounted for if deep water cycling and strong subduction zones are considered in tandem. Subduction zones that are so strong that they provide the dominant resistance to plate motion are not statistically preferred (that is, mantle viscosity remains an important variable for determining plate velocities). A change in tectonic regime is not required to account for the petrologic data trends. \\

There are added model implications that can provide enhanced layers of hypothesis testing. The cases from our modeling suite, that can match thermal constraints, suggest a change in the balance of mantle de- and re-watering at 2.0-2.5 Gya as subduction zones became cool enough to recycle a level of water that could keep pace with or outweigh mantle de-watering. This implies a switch from relatively dry to wet subduction, here caused by a thickening of the hydrated layer in subducting plates as the Earth cooled. An increase in water volume entering into the mantle at subduction zones can result in the production of felsic rather than mafic crust. As the amount of Earth's surface area covered by felsic crust increases, its oxidative efficiency decreases, leading to a rise in atmospheric O$_2$ [Lee et al., 2016]. A rise in atmospheric O$_2$ should then be coincident with the onset of the Earth's most recent cooling phase as constrained by petrological data [Condie et al., 2016; Herzberg et al., 2010]. Available data constraints [Lee et. al., 2016] are consistent with this model implication.  \\

In order for our models to match present day Urey ratio values, mantle re-watering needs to be in balance with or exceed de-watering from the termination of a flat line cooling phase to the present day (i.e., from 2.0 Gya to the present). Models in which re-watering is exceeding mantle de-watering over this time scale lead to lower Urey ratio values [Crowley, 2011] putting them deeper into the allowable data constraint range from Jaupart et al. [2007]. The balance of mantle de-watering and re-watering over geologic time is difficult to constrain. Efforts to do so do, however, suggest that mantle re-watering has exceeded de-watering over the last 500-600 Million years [Rupke et al., 2004; Parai and Mukhopadhyay, 2012]. This is consistent with the water balance  implications from our model cases that can match thermal history constraints. \\

As a final word, we offer a caution on extrapolating our results to other terrestrial planets. Although observations can constrain the range of viable model solutions for the Earth's thermal evolution, this does not mean that the same range needs to hold for terrestrial planets in general - be they planets in this solar system or in others. Theoretical models applied over geologic time frames need to effectively tune a range of parameter uncertainties by using Earth constraints. Model cases that fall out of the Earth viable range are not considered physically implausible. They are, instead, potential model paths for a terrestrial planet's evolution that are not in line with constraints on the evolution of a particular terrestrial planet (Earth). Thermo-tectonic history models have been and continue to be extrapolated to terrestrial planets orbiting stars other that our own. Often this is done by adjusting only a few variables, e.g., planetary size, while leaving others constant. The diversity of solutions for models that couple water cycling to planetary thermal evolution (Figures 3 and 4) highlight a deficiency in this approach and argue for a shift toward a fully statistical/probabalistic approach that does not hinge on Earth tuned results. \\


\section*{References}

\noindent
Christensen, U.R., 1984. Heat transport by variable viscosity convection and
implications for the Earth’s thermal evolution. Phys. Earth Planet. Int. 35,
264–282. \\

\noindent
Christensen, U.R., 1985. Thermal evolution models for the earth. J. Geophys. Res. 90
(2995–3007), B4. \\

\noindent
Condie, K., Aster, R.C., and van Hunen, J., 2016, A great thermal divergence in the mantle beginning 2.5 Ga: Geochemical constraints from greenstone basalts and komatiites, Geoscience Frontiers, 7, 543-553.  \\

\noindent
Conrad, C. and Hager, B., 1999, The thermal evolution of an earth with strong subduction zones, Geophysical Research Letters, 26, 3041-3044. \\

\noindent
Crowley, J.W., Gerault, M., and O'Connell, R.J., 2011, On the relative influence of heat and water transport on planetary dynamics, Earth Planet. Sci. Lett., 292, 79-88. \\

\noindent
Davies, G., 1980, Thermal histories of convective earth models and constraints on radiogenic heat production of earth, J. Geophys. Res., 85, 2517-2530. \\

\noindent
Grigne, C., Labrosse, S., 2001. Effects of continents on Earth cooling: thermal blanketing and depletion in radioactive elements. Geophys. Res. Lett. 28, 2707– 2710. \\

\noindent
Herzberg, C., Condie, K., and Korenaga, J., 2010, Thermal history of the Earth and its petrological expression, Earth Planet. Sci. Lett., 310, 380-388. \\

\noindent
Hirschmann, M. M., 2000, Mantle solidus: Experimental constraints and the effects of peridotite composition, Geochem. Geophys. Geosyst., 1(10), 1042, doi:10.1029/2000GC000070. \\

\noindent
Hoink, T., Lenardic, A. and Jellinek A. M., 2013, Earth's thermal evolution with multiple convection modes: A Monte-Carlo approach, Phys. Earth Planet. Inter., 221, 22-26. \\

\noindent
Jaupart, C., Labrosse, S., Mareschal, J., 2007. Temperatures, heat and energy in the mantle of the earth, treatise on geophysics. Mantle Dynam. 7, 253–303. \\

\noindent
Lenardic, A., Cooper, C., Moresi, L., 2011. A note on continents and the Earth’s Urey ratio. Phys. Earth Planet. Int. 188, 127–130. \\

\noindent
Katz, R.F., Spiegelman, M. and Langmuir, C.H., 2003, A new parameterization of hydrous mantle melting, Geochem. Geophys. Geosyst., 4(9), 1073. \\

\noindent
Kohlstedt, D. L. 2006, The role of water in high-temperature rock deformation, Rev. Mineral. Geochem., 62, 377–396, doi:10.2138/rmg.2006. 62.16. \\

\noindent
Korenaga, J., 2003, Energetics of mantle convection and the fate of fossil heat, Geophysical Research Letters, 30, 1437-1440. \\

\noindent
Korenaga, J., 2016, Can mantle convection be self-regulated?, Sci. Adv., 2016, 2:e1601168. \\

\noindent
Lee, C. A., Yeung, L. A., McKenzie, N. R., Yokoyama, Y., Ozaki, K. and Lenardic, A., 2016, Two-step rise of atmospheric oxygen linked to the growth of continents, Nat. Geosci., 9, 417-424.   \\

\noindent
Li, Z.-X. A., Lee, C.-T. A., Peslier, A.H., Lenardic, A. and Mackwell, S.J., 2008, Water contents in mantle xenoliths from the Colorado Plateau and vicinity: Implications for the mantle rheology and hydration-induced thinning of continental lithosphere, J. Geophys. Res., 113, B09210. \\

\noindent
McNamara, A. K., and P. E. van Keken (2000), Cooling of the Earth: A parameterized convection study of whole versuslayered models, Geochem. Geophys. Geosyst., 1, 1027, doi:10.1029/2000GC000045. \\

\noindent
Moore, W.B., and A. Lenardic, 2015, The efficiency of plate tectonics and non-equilibrium dynamic evolution of planetary mantles, Geophys. Res. Lett.,42, doi:10.1002/2015GL065621, 2015. \\

\noindent
Parai, R., and Mukhopadhyay, S., 2012, Water in the oceanic upper mantle: Implications for rheology, melt extraction and the evolution of the litho- sphere, Earth Planet. Sci. Lett., 317-318, 396–406. \\


\noindent
Rupke, L. H., J. P. Morgan, M. Hort, and J. A. D. Connolly, 2004, Serpen- tine and the subduction zone water cycle, Earth Planet. Sci. Lett., 223, 17–34, doi:10.1016/j.epsl.2004.04.018. \\

\noindent
Sandu, C., Lenardic, A. and McGovern, P.J., 2011, The effects of deep water cycling on planetary thermal evolution, Journal of Geophysical Research, 116, B12404. \\

\noindent
Schubert, G., Stevenson, D., Cassen, P., 1980. Whole planet cooling and the radiogenic heat source contents of the earth and moon. J. Geophys. Res. 85, 2531–2538. \\

\noindent
Ulmer, P., and V. Trommsdorff, 1995, Serpentine stability to mantle depths and subduction-related magmatism, Science, 268(5212), 858–861, doi:10.1126/science.268.5212.858.

\clearpage

\newpage 
 
\section{Appendix: Methods}

\subsection{Convection}
\indent Here we use the one-dimensional energy equation [Schubert et. al., 1979; Schubert et. al., 1980] to obtain the time rate of change of the Earth's average mantle temperature ($\dot{T}$)
\begin{equation} \label{PT_energy_balance}
\rho CV\dot{T} = -3 Aq_m + VQ(t)
\end{equation}
where $\rho$ is mantle density, $C$ is mantle heat capacity, and $q_m$ is the mantle surface heat flux. Mantle volume, $V$ and surface area $A$ are calculated as $R_m^3-R_c^3$ and $R_m^2-R_c^2$, respectively, where $R_m$ is the radius of the mantle and $R_c$ is the radius of the core. In this equation $T$ is the spherically averaged mantle temperature sense. In (\ref{PT_energy_balance}) it is assumed that the system is internally heated with an insulating bottom boundary. Heat is produced by radiogenic decay according to
\begin{equation} \label{heat_src}
Q(t) = Q_0e^{-\lambda t}
\end{equation}
where $Q_0$ and $\lambda$ are constants and $t$ is the current time.\\
\indent The Urey ratio $(Ur)$, which defines the ratio of heat produced within to heat transferred through the mantle surface, is defined as 
\begin{equation} \label{Ur}
Ur = \frac{VQ}{Aq_m}.
\end{equation}
When $Ur>1$, the planetary mantle is heating up. Alternatively, a value of $Ur<1$ means heat flow out of the mantle exceeds heat generated within the mantle and thus the mantle cools. \\
\indent The Rayleigh number $\left(Ra\right)$, a ratio of forces driving convection to those resisting it, is defined as
\begin{equation} \label{Ra}
Ra = \frac{g\alpha \Delta TZ^3}{\eta\kappa}
\end{equation} 
where $g$, $\alpha$, $Z$, $\eta$ and $\kappa$ are gravity, thermal expansivity, depth of the convecting layer, kinematic viscosity and thermal diffusivity, respectively. The value $\Delta T$ is the temperature difference driving convection defined as $T - T_s$, the difference between the mantle and surface temperatures. The relationship between a nondimensional heat flux ($Nu$) and $Ra$, which takes the form 
\begin{equation} \label{Nu}
Nu = \frac{q_mZ}{k\Delta T} = \left(\frac{Ra}{Ra_{cr}}\right)^\beta
\end{equation}
is used to solve for $q_m$, where $k$ is thermal diffusivity, $Ra_{cr}$ is the critical Rayleigh number which determines the onset of convection and $\beta$ is a scaling exponent. The value of $\beta$ for classic thermal history models is assumed to be 1/3 [Turcotte et. al., 1967; Solomatov 1995]. This assumes a plate tectonics mode of behavior in which the dominant resistance to plate motion comes from mantle viscosity. Lower values are also considered to mimic the effects of enhanced resistance coming from the strength of plates and/or plate margins [Christensen, 1985; Conrad and Hager, 1999; Korenaga, 2008]. \\
\indent A velocity scale $(u_c)$ is needed to compute degassing and regassing of the mantle. Fourier's law is rearranged to determine the lithospheric thickness according to the equation 
\begin{equation} \label{Db}
D_b = k\frac{\left(T_b - T_s\right)}{q_m}
\end{equation} 
where \emph{k} is thermal conductivity. The temperature $T_b$ represents the temperature at the base of the lithosphere and is equivalent to $T_m$. We use boundary layer theory [Schubert et. al., 2001] to derive a boundary layer breakaway time associated with subduction of the lithosphere according to 
\begin{equation} \label{ts}
t_s = \frac{1}{5.38\kappa_m}D_b^2
\end{equation}
From here, an equation for convective velocity is expressed as
\begin{equation} \label{uc}
u_c = \frac{\left(R_m - R_c\right)}{t_s}.
\end{equation}
The convective velocity scaling is of the form of $u_c \sim Ra^{2\beta}$, or more fully
\begin{equation} \label{uc_Ra}
u_c = \frac{a_1\kappa}{2\left(R_m-R_c\right)}\left(\frac{Ra}{Ra_{crit}}\right)^{2\beta}
\end{equation} 
where $a_1$ is a scaling parameter. It has a value  of 5.38 for the classic case of $\beta = 1/3$ [Schubert et. al., 1979; Schubert et. al., 1980]. As can be seen in (\ref{uc_Ra}), velocity has a power dependence on $\beta$. In the endmember case that $\beta = 0$, a constant velocity would be maintained. That is to say that for any value of $Ra$, the velocity will not change. To account for variable $\beta$, in a manner that allows all model to match present day plate velocities, a velocity scale must be set. To set this scale, we  use the present day values of convective vigor and velocity, $Ra_{now}$ and $u_{now}$, respectively, with the $\beta = 1/3$  scaling used as a reference. This leads to
\begin{align}
u_{now} &= a_1\frac{\kappa}{2\left(R_m-R_c\right)}\left(\frac{Ra_{now}}{Ra_{crit}}\right)^{\frac{2}{3}} \\
u_{now} &= a_2\frac{\kappa}{2\left(R_m-R_c\right)}\left(\frac{Ra_{now}}{Ra_{crit}}\right)^{2\beta} \\
a_2 &= a_1Ra_{crit}^{2\beta-\frac{2}{3}}Ra_{now}^{\frac{2}{3}-2\beta}
\end{align}
This value of $a_2$ can be computed for each $\beta$ and provide calibrated velocity scalings that will result in comparable present day velocity for model suites with variable $\beta$ values.  

\subsection{Volatile Cycling}
\indent Volatile cycling between the interior and surface has a direct influence on thermal evolution. The calculation of volatile cycling follows Sandu et. al. [2011]. Water leaves the mantle as an incompatible element participating in the batch melting process, which only occurs at mid-ocean ridges in our simplified model. Water is returned to the mantle via subduction processes.\\

To track mantle melting, the average mantle temperature calculated from (\ref{PT_energy_balance}) is converted to a temperature versus depth profile consisting of two parts: the conductive, lithospheric profile and the adiabat from the convecting mantle. The near surface temperature gradient is defined by the temperature at the surface and heat flux at the base of the lithosphere by
\begin{equation} \label{Lith_temp}
T(z)|_{z\leq D_b} = T_s + \frac{q_m}{k}z.
\end{equation} 
The adiabat contribution to the thermal profile is calcuated by converting the average mantle temperature to a potential temperature and projecting to depth according to 
\begin{equation} \label{adiabatic_temp}
T(z)|_{z>D_b} = T_p + \frac{g\alpha T_m}{C_p}z.
\end{equation} 
Although a depth dependent temperature profile is being calculated, it does not influence the convective dynamics. \\
\indent The thermal profile is compared to a solidus to compute the amount of melt generated by upwelling mantle. Two second-order polynomial functions are used to track melt fraction [Hirschman, 2000]. The solidus defines the temperature depth profile below which all mantle material will remain in its solid phase. As temperature warms beyond the solidus, a greater fraction of the mantle will melt until the liquidus, the temperature at which the entire mantle parcel becomes melted, is reached. In the case of hydrous melting, these functions take the form
\begin{equation} \label{Solidus}
T_{sol-hydr} = T_{sol-dry}-\Delta T_{H_20}
\end{equation} 
\begin{equation} \label{liquidus}
T_{liq-hydr} = T_{liq-dry}-\Delta T_{H_20}
\end{equation} 
where $T_{sol-dry}$, $T_{liq-dry}$, $T_{sol-hydr}$ and $T_{liq-hydr}$ are the dry solidus and liquidus and hydrated soldus and liquidus, respectively. The last term of both equations is the temperature shift  of each curve brought on by consideration of hydrous melting. This adjustment temperature scales with water concentration in the melt according to 
\begin{equation} \label{dT_H2O}
\Delta T_{H_20} = KX_{melt}^\gamma
\end{equation} 
where $K$ and $\gamma$ are constants which were calibrated by Katz et. al. [2003]. The parameter $X_{melt}$ is the ratio of water in the melt fraction expressed in kg of water per kg of melt and is calculated as 
\begin{equation} \label{X_melt}
X_{melt} = \frac{C_{mv}}{D_{H_2O}+F_{melt}\left(1-D_{H_2O}\right)}
\end{equation} 
where $C_{mv}$, $D_{H_2O}$ and $F_{melt}$ are the bulk water composition in the solid mantle expressed as a weight fraction, the bulk distribution coefficient which takes the value of 0.01 -- highlighting it behaves as an incompatible trace element -- and the degree of melting expressed as melt fraction, respectively. The melt fraction is parameterized in power-law form as
\begin{equation} \label{F_melt}
F_{melt} = \frac{T-\left(T_{sol-dry}-\Delta T_{H_2O}\left(X_{melt}\right)\right)}{T_{liq-dry}-T_{sol-dry}}^\beta .
\end{equation} 
This definition of $F_{melt}$ is valid from the surface to a depth of 300 km as constrained by observation and melting experiments. The values of melt fraction and water concentration were integrated over the melt zone thickness to provide an average to be used in the water budget calculation. \\

\begin{table*}[t!]
\caption{Model Parameters} 
\centering 
\begin{tabular}{l c c c} 
\hline \\ 
Parameter Name & Symbol & Value & Unites \\ [0.5ex] 
\hline \\ 
Initial mantle temperature & $T_{mi}$ & 1300, 2300, 3300 & K \\ 
Initial surface temperature & $T_{si}$ & 300 & K \\
\\
Initial radioactive heat & $Q_{mi}$ & 4.51, 3.157 & J/(m$^3$yr) \\
Radioactive decay constant & $\lambda$ & 3.4*10$^{-10}$ & yr$^{-1}$ \\
\\
Lower mantle boundary (Earth) & $R_c$ & 3471 & km \\
Upper mantle boundary (Earth) & $R_m$ & 6271 & km \\
Initial lithospheric thickness & $Z_{lith}$ & 0 & km \\
\\
Mantle density & $\rho$ & 3000 & kg/m$^3$ \\
Thermal conductivity & $k$ & 4.2 & W/(m*K) \\
Specific heat & $C_p$ & 1400 & J/(kg*K) \\
Coefficient of thermal expansion & $\alpha$ & 3*10$^{-5}$ & K$^{-1}$ \\
\\
Viscosity constant & $\eta_0$ & 1.7*10$^{17}$ & Pa*s \\
Viscosity material constant & $Acre$ & 90 & MPa$^{-r}$/s \\
Viscosity exponent constant & $r$ & 1.2 & - \\
Activation energy for creep & $AE_c$ & 4.8*$^5$ & J/mol \\ 
\\
Degassing efficiency factor & $\chi_d$ & 0.002, 0.02, 0.04, 0.4 & - \\
Regassing efficiency factor & $\chi_r$ & 0.001, 0.003, 0.01, 0.1 & - \\
Initial Mantle Ocean Masses & $OM_i$ & 0.1, 0.25, 0.5, 0.75, 1, 2, 4, 6 & - \\
\\
Fraction of volatile in basalt & $f_{bas}$ & 0.03 & - \\
Density of basalt & $\rho_{basalt}$ & 2950 & kg/m$^3$ \\
Average thickness of basalt & $Z_{basalt}$ & 5000 & m \\
\\
Critical Rayleigh number & $Ra_{crit}$ & 1100 & - \\ 
Nusselt-Rayleigh scaling parameter & $\beta$ & 0.1, 0.15, 0.2, 0.25, 0.3, 0.33 & - \\
[1ex] 
\hline 
\end{tabular}
\label{table:Params} 
\end{table*}

\indent The melt zone thickness is dependent upon the relative positioning of the thermal profile and the solidus. The lower boundary of the melt zone is defined where a parcel of upwelling mantle reaches a temperature hot enough to begin producing partial melt. This depth will be identified by the intersection between the mantle thermal profile and solidus. The upper bound of the melt zone is defined by the near surface intersection of the solidus and thermal profile. At this depth, the upwelling mantle has cooled enough such that no more melt is being produced. The depth difference between these two cross over points defines the thickness of the mantle undergoing partial melting and contributes to the mantle degassing calculation. \\

Water is degassed at mid-ocean ridges (MOR). The rate at which water is degassed depends on the volume of mantle transiting the melt zone below the ridge, the amount of melt produces, what fraction of that melt is water and how much of that water makes it to the surface. In equation form, the degassing rate ($r_{MOR}$) is 
\begin{equation} \label{rMOR}
r_{MOR} = \rho_mF_{melt}X_{melt}D_{melt}S\chi_d
\end{equation}
where $F_{melt}$ is the integrated melt fraction in the melt zone and $\chi_d$ is the degassing efficiency factor. Both $D_{melt}$ and $X_{melt}$ are calculated according to the parameterization of Katz et. al. [2003]. The areal spreading rate, $S$, is derived from a boundary layer model [Schubert et. al., 2001] and is represented as
\begin{equation} \label{areal_spreading}
S = 2L_{ridge}u_c
\end{equation}
which assumes symmetrical spreading along a constant length of ridge, $L_{ridge}$. Velocity, $u_c$, is solved within the model according to equation (\ref{uc}).\\

Water is assumed to be returned to the mantle at subduction zones. This water is bound in the serpentinized and thin sedimentary layers of the slab [Rupke et. al., 2004]. Since most water held in the sedimentary layer is lost back to the surface, the serpentinized layer is the more important factor in our calculation. The rate at which water is subducted back into the mantle ($r_{SUB}$) is
\begin{equation} \label{rSUB}
r_{SUB} = f_h\rho D_{hydr}S\chi_r,
\end{equation}
where $f_h$, $D_{hydr}$, and $\chi_r$ are the mass fraction of volatiles in the serpentinized layer, thickness of the serpentinized layer and regassing efficiency factor, respectively. In this case, $D_{hydr}$ is defined as the depth of the 700$^o$C isotherm as the hydrous phase of serpentine decomposes around this temperature [Ulmer  and Trommsdorff, 1995]. \\

Individually, equations (\ref{rMOR}) and (\ref{rSUB}) tell what is occurring independently. The overall flow rate of mantle water ($r_{Mmv}$) is
\begin{equation} \label{water_balance}
r_{Mmv} = r_{SUB}-r_{MOR}.
\end{equation}
This tracks the balance of water between the interior and surface reservoirs. When $r_{Mmv}$ is positive, the mantle is being replenished with water. When it is negative, the mantle is losing water. If $r_{Mmv}$ is zero, there is a balance between the two and the mass of water in each reservoir remains constant.

\subsection{Viscosity}
\indent The temperature dependence of mantle viscosity takes the Arrenhius form
\begin{equation} \label{nv_visc}
\eta = \eta_0exp\left(\frac{A}{RT_m}\right)
\end{equation}
where $\eta_0$, $A$, $R$ are a reference viscosity, activation energy for creep  (Weertman and Weertman, 1975) and the universal gas constant. \\
\indent Experiments have shown hydration effects have a power law effect on mantle viscosity [Carter et. al., 1970; Chopra and Paterson, 1984; Mackwell et. al., 1985; Karato and Wu, 1993] The power law was further refined to include dependence on water fugacity in olivine [Hirth and Kohlstedt, 1996; Mei and Kohlstedt, 2000]. Assuming Newtonian behavior and an empirical relation for water fugacity based on concentrations [Li et. al., 2008], the effective viscosity is given by
\begin{align} \label{vd_visc}
\eta_{eff} &= \frac{\tau}{\dot{\epsilon}} = \eta_0 A_{cre}^{-1}\left(exp\left(c_0 + c_1 ln C_{OH} + c_2 ln^2 C_{OH} \right. \right. \nonumber \\
&\left. \left. + c_3ln^3 C_{OH}\right)\right)^{-r} exp\left(\frac{A}{RT}\right)
\end{align}
where $\tau$, $\dot{\epsilon}$ are stress and strain rate. Experimentally determined constants from Li et al. [2008] are $c_0$, $c_1$, $c_2$ and $c_3$ and $C_{OH}$ is the water concentration expressed as $H/10^6$ Si. Here $\eta_0$ and $A_{cre}$ are a calibration and material constant.  

\section{Appendix: Model Output Statistics}

\begin{figure}[b!]
  \centering
  \includegraphics[width=\linewidth]{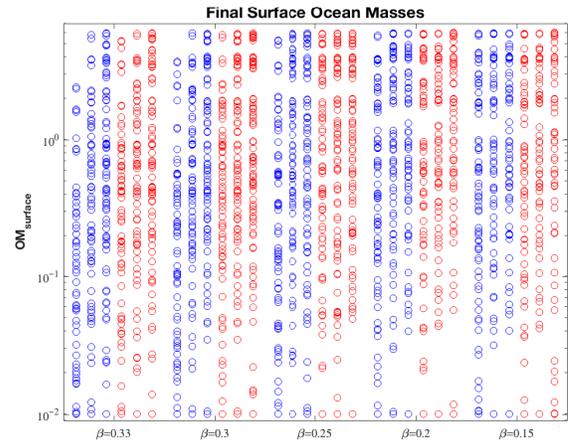}
  \caption{Each open circle represents the final present day amount of water at the Earth's surface for a specific combinations of parameters and initial conditions. The final distribution of water is sensitive to the assumed initial amount of water in the system. This initial condition sensitivity made the final water distribution a weak discriminate between different model hypotheses. That is, changes in the assumed initial water volume allowed the full range of model hypothesis to match final surface water distribution with a weak statistical preference between hypotheses.}
\end{figure}

\begin{table*}[t!]
\caption{Number of Models Successfully Constrained} 
\centering 
\begin{tabular}{c|cccccc|c} 
\hline 
 & $\beta=0.33$ & $\beta=0.3$ & $\beta=0.25$ & $\beta=0.2$ & $\beta=0.15$ & $\beta=0.1$ &  \\ 
\hline  
LQLT & 0  & 0  & 0  & 0 & 0 & 0 & 0 \\
LQMT & 41 & 29 & 16 & 5 & 0 & 0 & 91 \\
LQHT & 35 & 3  & 7  & 0 & 0 & 0 & 45 \\
HQLT & 0  & 5  & 0  & 3 & 0 & 0 & 8 \\
HQMT & 0  & 3  & 10 & 0 & 0 & 0 & 13 \\
HQHT & 2  & 3  & 10 & 0 & 0 & 0 & 15 \\
\hline 
     & 78 & 43 & 43 & 8 & 0 & 0 & 172 \\
\hline 

\end{tabular}
\end{table*}

Table 1 shows the range of model parameters, and Table 2 shows the models that can match data constraints broken into model subsets by variable $\beta$, initial heat source density, and initial temperatures. Of all the model cases, 3.7 percent can match data constraints. None of the cases with a $\beta$ value below 0.2 can match all the constraints. Once those cases are removed, 5.6 percent of the remaining cases can match data constraints. Statistically, models with initial heat source densities lower than chondritic [Jackson and Jellinek, 2013] are preferred. Once the higher heat source density cases are removed, 8.9 percent of the remaining cases can match data constraints. Once the coolest initial condition cases, which are not statistically preferred, are removed, 13.3 percent of the remaining cases can match data constraints. The number of cases within that reduced set is 1024. Model cases with low initial mantle water have a statistically lower chance of allowing for a mantle dewatering effect that can drive a flat line cooling phase and associated delayed mantle cooling (Figure 2). If models are further limitted to an initial mantle water content of at least one ocean mass equivalent, then 19.14 percent of the remaining models (512 total) can match data constraints. 

Figure 5 shows final water distributions. The sensativity of final surface water on assumed initial mantle water volume made this a weak descriminate between different model hypotheses. It should be further noted that the surface water in our models represents water that is degassed from the mantle. Late stage water delivery could alter the final surface volume of water without having a significant effect on our thermal evolution models provided the initial mantle water is not significantly lower than one ocean mass (it was only the lower initial mantle water cases that allowed mantle re-watering to be limitted due to the lack of surface water). This further weakens the use of the final surface water as a constraint that can descriminate between competing hypothesis regarding thermal evolution. 

\section*{Appendix: References}

\noindent
Carter, N.L., and Ave’Lallemant, H.G., 1970, High temperature flow of dunite and peridotite, Geol. Soc. Am. Bull., 81, 2181–2202. \\

\noindent
Chopra, P. N., and M. S. Paterson (1984), The role of water in the deformation of dunite, J. Geophys. Res., 89, 7861–7876. \\

\noindent
Christensen, U., 1985, Thermal evolution models for the earth, Journal of Geophysical Research, 90, 2995-3007. \\

\noindent
Conrad, C. and Hager, B., 1999, The thermal evolution of an earth with strong subduction zones, Geophysical Research Letters, 26, 3041-3044. \\

\noindent
Hirschmann, M. M., 2006, Water, melting, and the deep Earth H2O cycle, Annu. Rev. Earth Planet. Sci., 34, 629–653. \\

\noindent
Hirth, G., and Kohlstedt, D.L., 1996, Water in the oceanic upper mantle: Implications for rheology, melt extraction and the evolution of the litho- sphere, Earth Planet. Sci. Lett., 144, 93–108. \\

\noindent
Karato, S., and Wu, P., 1993, Rheology of the upper mantle: A synthesis, Science, 260, 771–778. \\

\noindent
Katz, R.F., Spiegelman, M. and Langmuir, C.H., 2003, A new parameter- ization of hydrous mantle melting, Geochem. Geophys. Geosyst., 4(9), 1073. \\

\noindent
Korenaga, J., 2008, Urey ratio and the structure and evolution of earth's mantle, Reviews of geophysics, 46. \\

\noindent
Li, Z.-X. A., Lee, C.-T. A., Peslier, A.H., Lenardic, A. and Mackwell, S.J., 2008, Water contents in mantle xenoliths from the Colorado Plateau and vicinity: Implications for the mantle rheology and hydration-induced thinning of continental lithosphere, J. Geophys. Res., 113, B09210. \\

\noindent
Mackwell, S.J., Kohlstedt, D.L. and Paterson, M.S., 1985, The role of water in the deformation of olivine single crystals, J. Geophys. Res., 90, 11319–11333. \\

\noindent
Mei, S., and D. L. Kohlstedt, 2000, Influence of water on plastic deformation of olivine aggregates 2. Dislocation creep regime, J. Geophys. Res., 105, 21,471–21,481. \\

\noindent
Rupke, L.H., Morgan, J.P., Hort, M. and Connolly, J.A.D., 2004, Serpentine and the subduction zone water cycle, Earth Planet. Sci. Lett., 223, 17–34. \\

\noindent
Sandu, C., Lenardic, A. and McGovern, P.J., 2011, The effects of deep water cycling on planetary thermal evolution, Journal of Geophysical Research, 116, B12404. \\

\noindent
Schubert, G., P. Cassen, and R. E. Young, 1979, Subsolidus convective cooling histories of terrestrial planets, Icarus, 38, 192–211. \\

\noindent
Schubert, G., D. Stevenson, and P. Cassen, 1980, Whole planet cooling and the radiogenic heat source contents of the Earth and Moon, J. Geophys. Res., 85, 2531–2538. \\

\noindent
Schubert, G., D. L. Turcotte, and P. Olson, 2001, Boundary layer theory, in Mantle Convection in the Earth and Planets, pp. 350–361, Cambridge Univ. Press, Cambridge, U. K. \\

\noindent
Solomatov, V.S., 1995, Scaling of temperature- and stress-dependent viscosity convection, Phys. Fluids, 7, 266-274. \\

\noindent
Turcotte, D.L. and Oxburgh, E.R. Finite amplitude convective cells and continental drift, J. Fluid Mech. 28 (1967) 29–42. \\

\noindent
Ulmer, P. and Trommsdorff, V., 1995, Serpentine stability to mantle depths and subduction-related magmatism, Science, 268(5212), 858–861. \\

\noindent
Weertman, J. and Weertman, J.R., 1975, High temperature creep of rock and mantle viscosity. Annual Review of the Earth and Planetary Sciences 3: 293–315. \\

\newpage



\end{document}